\documentclass[twocolumn,showpacs,preprintnumbers,amsmath,amssymb,
superscriptaddress]{revtex4}
\usepackage[dvips]{color}

\usepackage{graphicx}% Include figure files
\usepackage{dcolumn}% Align table columns on decimal point
\usepackage{bm}% bold math
\usepackage{dsfont}% nice one operators

\newcommand{\inistat}{\bm{\zeta}_{\frac{\pi}{2}}}
\newcommand{\sn}{\operatorname{sn}} \newcommand{\cn}{\operatorname{cn}}
\newcommand{\dn}{\operatorname{dn}} \newcommand{\degree}{^\circ}

\begin{document}

\preprint{APS/123-QED}
\title{Evolution of a spinor condensate: coherent dynamics, dephasing
  and revivals}% Force line breaks with \\

\author{J. Kronj{\"a}ger}
\affiliation{Institut f{\"u}r Laserphysik, Universit{\"a}t
Hamburg, Luruper Chaussee 149, D-22761 Hamburg, Germany.}
\author{C. Becker}
\affiliation{Institut f{\"u}r Laserphysik, Universit{\"a}t
Hamburg, Luruper Chaussee 149, D-22761 Hamburg, Germany.}
\author{M. Brinkmann}
\affiliation{Institut f{\"u}r Laserphysik, Universit{\"a}t
Hamburg, Luruper Chaussee 149, D-22761 Hamburg, Germany.}
\author{R. Walser}
\affiliation{Abteilung Quantenphysik, Universit\"{a}t Ulm,
D-89069 Ulm, Germany}
\author{P. Navez}%
\affiliation{Labo Vaste-Stoffysica en Magnetisme, Katholieke
Universiteit Leuven, Celestijnenlaan 200D, B-3001 Leuven, Belgium.}%
\affiliation{present address: Universit{\"a}t Duisburg-Essen,
Universit{\"a}tsstrasse 5, 45117 Essen, Germany}
\author{K. Bongs}
\affiliation{Institut f{\"u}r Laserphysik, Universit{\"a}t
Hamburg, Luruper Chaussee 149, D-22761 Hamburg, Germany.}
\author{K. Sengstock}
\affiliation{Institut f{\"u}r Laserphysik, Universit{\"a}t Hamburg,
  Luruper Chaussee 149, D-22761 Hamburg, Germany.}

\date{\today}% It is always \today, today,
\begin{abstract}
  We present measurements and a theoretical model for the interplay of spin
  dependent interactions and external magnetic fields in atomic spinor
  condensates. We highlight general features like quadratic Zeeman dephasing
  and its influence on coherent spin mixing processes by focusing on a
  specific coherent superposition state in a $F=1$ $^{87}$Rb Bose-Einstein
  condensate.  In particular, we observe the transition from coherent spinor
  oscillations to thermal equilibration.
\end{abstract}

\maketitle

Multi-component Bose-Einstein condensates with spin degree of freedom,
the so called spinor condensates, are experiencing rapidly growing
attention.  These ultra-cold quantum gas systems are interesting in
several respects, e.g.~they show intriguing static and dynamic
magnetic properties~\cite{Ho1998a,Koashi2000a,Stamper-Kurn2001a}, they
represent a well controlled thermodynamic model
system~\cite{Lewandowski2003a,Erhard2004a} and they promise unique
insights into mesoscopic multi-component entanglement and decoherence
processes~\cite{Sorensen2001a}.

Pioneering experiments on $F=1$ $^{23}$Na spinor condensates,
found to be anti-ferromagnetic, and quasi-spin-1/2 systems in
$^{87}$Rb have shown fascinating demixing dynamics, metastability
and domain formation processes~\cite{Stamper-Kurn2001a,Hall1998a}.
Recently the spinor systems under study have been extended to
$^{87}$Rb, which behaves ferromagnetic in the $F=1$
state~\cite{Schmaljohann2004a,Chang2004a} and anti-ferromagnetic in
$F=2$~\cite{Schmaljohann2004a}. In particular, these experiments
also showed spinor oscillations and shifted the research interest
towards dynamic spin conversion effects. Recent studies have
demonstrated a magnetic field dependence of the oscillation
amplitude and
frequency~\cite{Kuwamoto2004a,Schmaljohann2004b,Widera2005a},
which was also theoretically assigned to an interesting resonance
phenomenon~\cite{Zhang2005a}.  The coherent evolution in these
experiments at short timescales is opposed to the incoherent
thermodynamic behavior at long timescales, e.g.  decoherence
driven cooling~\cite{Lewandowski2003a}, constant temperature
BEC~\cite{Erhard2004a} or temperature driven
magnetization~\cite{Schmaljohann2004b}.

The coherent and thermodynamic regime are conceptually very
different, as the first relies on the assumption of each atom in
the ensemble being in the same superposition state, while there
are presumably different and independent ensembles for each spin
state in the second regime, i.e. here an atom is either in states
$|F=1,m_F=\pm1\rangle$ or $|F=1,m_F=0\rangle$. It is still under
discussion, how and at which point the crossover from the ''pure
state'' ensemble to a ''mixed state'' arises and how this is
connected to decoherence and dephasing mechanisms in spinor
systems. This question can be generalized to the understanding of
decoherence in arbitrary multi-component systems and its
dependence on the number of constituents, which might have
important consequences for quantum information theory. So far
studies on the relative phases of different components have only
been performed in quasi-spin 1/2
systems~\cite{Hall1998b,Harber2002a,Lewandowski2003a} based on the
preparation of mixtures with well defined phase, while
investigations on spinor systems were essentially restricted to
population-based state preparation without phase control.

In this paper we discuss dephasing and decoherence effects in spinor quantum
systems. In particular we present measurements on the influence of finite
temperature and external magnetic fields on the evolution of a $F=1$ $^{87}$Rb
spinor condensate prepared in the $\bm{\zeta}_{\frac{\pi}{2}}=(\zeta_{+}, \zeta_{0},
\zeta_{-})^\top =(-1, -i\sqrt{2}, 1)^\top/2$ state.  In addition we develop an analytic
description for the mean-field evolution and emphasize the importance of the
quadratic Zeeman effect, which leads to periodic dephasing and rephasing of
the spin expectation value. We theoretically predict and experimentally find
dephasing induced spin dynamics with magnetic field dependent period and
amplitude, which clearly shows the coherence of spin mixing in spinor
condensates. An analysis of the data again independently confirms the $F=1$
state of $^{87}$Rb to be ferromagnetic.

\section{Theory}

In order to simplify the discussion and concentrate on the spinor physics, we
will restrict ourselves to the single mode approximation (SMA), which
effectively factorizes the $F=1$ spinor order parameter as
\begin{equation}
  \boldsymbol{\psi} (\bm{r},t) =
                                %\left(
                                %   \begin{array}{l}
  \begin{pmatrix}
    \psi_{+}(\bm{r},t) \\
    \psi_0 (\bm{r},t) \\
    \psi_{-}(\bm{r},t)
  \end{pmatrix}
                                %  \end{array} \right)
  =
  \sqrt{ n(\bm{r})}
  %\left(
  %  \begin{array}{c}
  \begin{pmatrix}
      \zeta_{+}(t) \\
      \zeta_0(t) \\
      \zeta_{-}(t)
   % \end{array}
  %\right),
    \end{pmatrix},
  \end{equation}
  where the volume density $n(\bm{r})$ and spinor $\bm{\zeta}$ are
  individually normalized to $\int d^3r\, n(\bm{r}) = N$ and
  $||\bm{\zeta}||^2=1$, respectively.  This assumption is justified,
  if the spin dynamics is slow compared to the timescale of motion in the trap and
  if the spin dependent interaction is much smaller than the spin independent
  one.

In SMA, the mean-field equations of motion for the spinor
are~\cite{Ho1998a,Ohmi1998a}
\begin{subequations}
  \label{eq:hamilton}
  \begin{eqnarray}
    \label{eq:hamiltona}
    i\,\partial_t\bm{\zeta}  & = & ({H}_\text{Z} +{H}_\text{mf})\, \bm{\zeta}, \\
    \label{eq:hamiltonb}
    {H}_\text{Z} & = & {H}_{\text{LZ}}+ {H}_{\text{QZ}}= -p\,
    {F}_z - q \,(\mathds{1}-{F}^2_z),\\
    \label{eq:hamiltonc}
    {H}_\text{mf} & = &g_2\, \langle n\rangle
    \sum_{\alpha=x,y,z} {F}_\alpha \,\bm{\zeta}\bm{\zeta}^\dagger{F}_\alpha,
  \end{eqnarray}
\end{subequations}
where $p \propto B$ and $q \propto B^2$ are the coefficients of
the linear and quadratic Zeeman effect respectively. For
$^{87}$Rb, F=1, the Breit-Rabi formula allows to calculate them as
$p\approx -\mu_B B/2 \hbar$ and $q \approx p^2/\omega_\text{hfs}$
with the $^{87}$Rb ground state hyperfine splitting
$\omega_{\text{hfs}} \approx 2\pi\cdot 6.835\, \mathrm{GHz}$. The
spin-dependent mean-field interaction is proportional to $g_2 =
4\pi\hbar/m\times (a_{F=2}-a_{F=0})/3$ \cite{Ho1998a} and $\langle
n\rangle =\int d^3r\,n^2(\bm{r})/N$. We will identify the
$3\times3$ matrices $\{{F}_x, {F}_y, {F}_z\}$ as the standard
representation of the $F=1$ angular momentum algebra
\cite{biedenharn}. In principle, the mean-field
Eqs.~(\ref{eq:hamilton}) could be derived also from a
Hamilton-Jacobi theory.  The conservation of spin and energy
follow naturally from rotational and time-translation symmetries.
Due to these constraints, it is indeed possible to find integrable
solutions to the full nonlinear equations of motion. In case of
the initial state $\inistat $, the time-dependent spinor
amplitudes can be written in terms of Jacobi elliptic
functions~\cite{Bronstein}
\begin{subequations}
  \label{eq:amplana}
  \begin{eqnarray}
    \zeta_0(t) & = & \frac{s}{\sqrt{2}}
    \left[ \frac{(1-k)\sn_k( \frac{qt}{2} )}{1-k\sn_k^2( \frac{qt}{2} )}
      - \frac{i\cn_k( \frac{qt}{2})\dn_k( \frac{qt}{2})}{1+
        k\sn_k^2( \frac{qt}{2} )}\right] \\
    \zeta_{\pm}(t) & = & \mp\frac{s \,e^{\pm ipt}}{2}
    \left[\frac{\cn_k( \frac{qt}{2} )\dn_k( \frac{qt}{2} )}{1-
        k\sn_k^2( \frac{qt}{2} )}
      - \frac{i(1+k)\sn_k( \frac{qt}{2})}{1+
        k\sn_k^2( \frac{qt}{2})} \right]\nonumber\\
  \end{eqnarray}
\end{subequations}
where $s=\exp(-i(g_2\langle n \rangle - q)t/2)$ is a dynamic global phase and
$k = g_2\langle n \rangle / q$ is the ratio of interaction energy to quadratic
Zeeman effect.  For the directly measurable populations, the solution
simplifies to
\begin{subequations}
  \label{eq:popana}
\begin{eqnarray}
  \label{eq:popanaa}
  |\zeta_0(t)|^2        & = & (1-k\sn^2_k(qt))/2,\\
  |\zeta_{\pm}(t)|^2  & = & (1+k\sn^2_k(qt))/4.
\end{eqnarray}
\end{subequations}
For small $|k| \ll 1$, the Jacobi elliptic functions can be
approximated by ordinary trigonometric ones, i.\thinspace{}e.,
$\sn_k(x)\approx\sin(x)$, $\cn_k(x)\approx\cos(x)$, $\dn_k(x)\approx 1$. Thus,
the populations of Eqs.~(\ref{eq:popana}) oscillate with an amplitude given by
$k$, and a period which is $\pi/q$ for small $|k|$.
This solution offers a clear way to determine the magnetic
properties of the system as $k$ is negative/positive for
ferromagnetic/anti-ferromagnetic systems, resulting in an
increase/decrease of $|\zeta_0|^2$ at the beginning of the
oscillation (if $q>0$).

At this point, we would like to emphasize the intricate connection
between quadratic Zeeman dephasing and spin mixing dynamics: for
zero quadratic Zeeman shift $q=0$, the state $\inistat$ does not
evolve according to Eq.~(\ref{eq:popana}), except for a trivial
Larmor precession around the magnetic field vector, which can be
compensated by choosing a rotating frame of reference. The
intuitive reason for this lies in the rotational symmetry of the
spin interactions, leading to a conservation of magnetization. In
a frame rotated by $90\degree$, the above state is just the
  fully stretched state ${\bm \zeta}=(0,0,1)$, which is an eigenstate of $H_\text{mf}$ and
  consequently does not evolve. In fact, any rotated stretched state belongs
  to this class of stationary states of $H_\text{mf}$ \footnote{Note that the
    class of rotated stretched states does not form an eigenspace of
    $H_\text{mf}$, since $H_\text{mf}$ is an effective Hamiltonian which in
    turn depends on the spin state.}.

  The quadratic Zeeman effect however destroys the fully
  stretched state, removing it from the class of stationary states,
  and thus allowing interaction driven spin dynamics.  Two regimes
can be distinguished in the resulting population oscillation: the {\it
  interaction dominated regime} at low magnetic fields, where the
oscillation period is essentially given by the intrinsic spin
interactions, and the {\it quadratic Zeeman regime}, where it is given
by the respective beat period $\pi/q$. In both limits $k\to 0$ and
$q\to 0$, the amplitude of population oscillations goes to zero.
Interestingly the crossover between the two regimes shows a
resonance-like structure, as also recently predicted for a more
general case~\cite{Zhang2005a}.

\section{experiment}

In our experimental setup described in~\cite{Schmaljohann2004a} we
initially prepare degenerate $^{87}$Rb ensembles containing up to
$10^6$ atoms in the $|F=1,m_F=-1\rangle$ state in an optical
potential with trapping frequencies $\omega_x : \omega_y :\omega_z
\approx 1:7:40$ and $\omega_x = 2\pi \cdot 16\ldots 25 \, {\rm
s}^{-1}$. Using a radio frequency $\pi/2$-pulse of typically
40$\mathrm{\mu s}$ duration, we evolve the spinor into the
superposition state $\bm{\zeta}(0) = \inistat$. In real space this
corresponds to a $90\degree$ rotation of the classical spin
expectation value $\boldsymbol{\mathcal{F}} = \bm{\zeta}^\dagger
\bm{F} \bm{\zeta} $ from $\boldsymbol{\mathcal{F}} = -e_z$ to
$\boldsymbol{\mathcal{F}} = e_y$.

\subsection{Phase evolution}
\begin{figure}[t]
\includegraphics{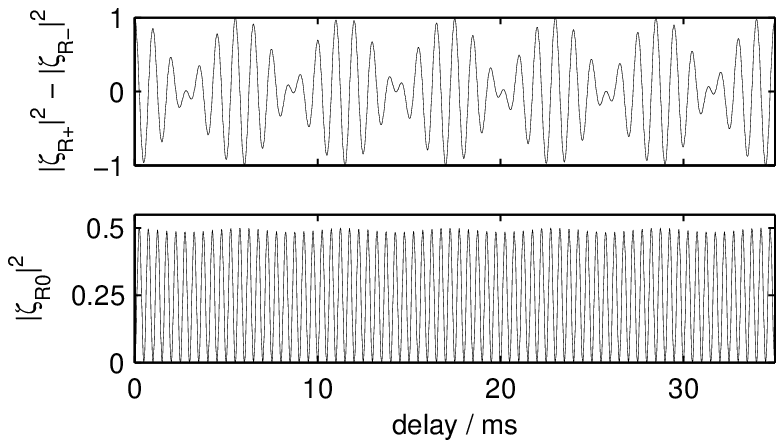}
\caption{\label{fig:Ramsey_th} { Theoretical prediction for the
  outcome of a Ramsey experiment with an interacting spin-1 spinor condensate,
  plotting the normalized magnetization in the direction of
  the quantization axis $|\zeta_{R+}|^2-|\zeta_{R-}|^2$ and the
  $m_F=0$-projection $|\zeta_{R0}|^2$ as a function of pulse delay time.  The
  parameters were chosen to be $|p|=2\pi \cdot 772\, $kHz (corresponding to
  $|q|= 2\pi \cdot 87.2\, $Hz), $k=-0.03$ and $\omega_{RF}=2\pi \cdot 773\, $kHz for the radio frequency
  field driving the $\pi /2 $-pulses.} }
\end{figure}

Using the classical picture it is evident, that another $\pi/2$
radio frequency pulse will rotate the spin vector upward to
$\boldsymbol{\mathcal{F}} = +e_z$ or equivalently the spinor to
$\bm{\zeta} = (1,0,0)^\top$, transferring all the population into
the $|F=1, m_F=+1\rangle$ state. The addition of a delay time in
between the two pulses results in the spin-1 analogon of the
famous Ramsey experiment for spin-1/2
particles~\cite{Ramsey1949a}, which is sensitive to the quantum
mechanical phase evolution of the system. The last $\pi/2$-pulse
creates superpositions of the spinor components and thus
transforms the phase information into population information
accessible to detection. Thus, the state of the Ramsey spinor is
${\bm \zeta}_{R}(t)=\exp{(-i \pi/2 \, F_x)}\, \exp{(i \omega_{RF}t
\, F_z)}\,{\bm \zeta }(t)$ with ${\bm \zeta }(t)$ from
Eqs.~\ref{eq:amplana}.

In the following we will analyze the time-dependent population
distribution in such a Ramsey experiment in order to gain insights
into dephasing and decoherence mechanisms.  The corresponding
analytical results in Fig.~\ref{fig:Ramsey_th}, clearly show a
magnetization oscillation with the detuning between the Larmor
precession frequency and the radio frequency for driving the
$\pi/2$-pulses, as one would also expect from spin-1/2 particles.

There is however an interesting "beat" feature with twice the
quadratic Zeeman frequency adding new physics to the spin-1
system. For the given parameters there are only very small
interaction effects barely visible as a modulation of the
$|\zeta_0|^2$-oscillation envelope and practically undetectable in
the magnetization.

In order to draw a simple picture of the corresponding physics, it
is thus sufficient to neglect interactions and to concentrate on
the evolution of a single spin-1 particle in an external magnetic
field aligned along the z-axis. In this picture the linear Zeeman
shift of Eq.~(\ref{eq:hamiltonb}) adds a time dependent phase of
equal magnitude but opposite in sign to the $\zeta_{+}$ and
$\zeta_{-}$ components: $\bm{\zeta}_{\text{LZ}}(t) =\exp{(-it
\,H_{\text{LZ}})}\,\inistat$$ =( -e^{-ipt}, -i\sqrt{2},
e^{ipt})^\top/2$. In real space this corresponds to a spin vector
$\boldsymbol{\mathcal{F}} = (\sin{(pt)}, \cos{(pt)}, 0)^\top$
rotating around the magnetic field axis, i.e.~the Larmor
precession of a magnetic moment, just as one would classically
expect.

A different ``non-classical'' behaviour arises from the quadratic
Zeeman effect, which corresponds to adding a time dependent phase
to the $\zeta_0$ component: $\bm{\zeta}_{\text{QZ}}(t) = \exp{(-it
\,H_{\text{QZ}})}\,\inistat$$ =(-1,
  -i \sqrt{2} e^{iqt},1)^\top/2$.  The resulting
spin expectation vector $\boldsymbol{\mathcal{F}}=( 0, \cos{(qt)},
0)^\top$ now periodically disappears and reappears, always
pointing along the original direction. In a Ramsey experiment, the
combined linear and quadratic Zeeman effect show up as a periodic
dephasing and rephasing of the rotating spin vector just as the
analytic result shown in Fig.~\ref{fig:Ramsey_th}, which can also
be interpreted as a beat note between the two slightly detuned
transitions $m_F=0 \leftrightarrow m_F=\pm 1$.

\begin{figure}[t]
\includegraphics{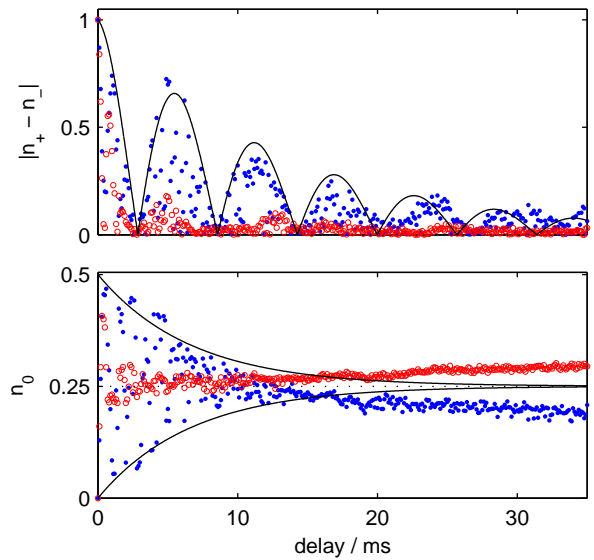}
\caption{\label{fig:Ramsey} Magnetization amplitude and $m_F=0$
  population after a Ramsey sequence as a function of the
  delay time between Ramsey pulses. The dots (blue) show the experimental
  results for the condensate fraction, while the circles (red)
  represent the normal component. The lines are phenomenological envelope
  fits with exponential decay to guide the eye. The magnetic field parameters
  are the same as in Fig.~\ref{fig:Ramsey_th}.}
\end{figure}

Fig.~\ref{fig:Ramsey} shows the experimental results of such a
Ramsey sequence consisting of a $\pi/2$-pulse for state
preparation, a variable delay time and another $\pi/2$-pulse
before Stern-Gerlach separation and imaging of the individual spin
components. In order to emphasize the effect of quadratic Zeeman
dephasing instead of the trivial Larmor precession, we plot the
magnitude of the magnetization vector instead of the magnetization
itself and concentrate on the envelope of the corresponding data
points~\footnote{We
  are able to follow the Larmor precession for 5\,ms before shot-to-shot phase
  fluctuations corresponding to a magnetic field noise of $\approx 200\mu
  $G lead to a random phase of the magnetization oscillation at the time of
  detection. This does not effect the detection of quadratic Zeeman dephasing
  of the coherent evolution, which can clearly be identified by looking at the
  envelope of the data points.}. In order to achieve the parameters used for
Fig.~\ref{fig:Ramsey_th} an offset field of $\approx 1.1 \,$G was
applied, corresponding to a Larmor precession frequency of
$|p|=2\pi\cdot 772 \,$kHz and a quadratic Zeeman beat frequency
  $|q|=2 \pi \cdot 87.2 $ Hz
  corresponding to a beat period of $\approx 5.7 \,$ms, i.e.~operating in the
  quadratic Zeeman regime. In this case already the single particle picture
discussed above clearly reproduces our experimental data: The
magnetization is modulated in amplitude by the quadratic Zeeman
effect with collapses and revivals corresponding to a beat period
of $\pi/q$.

The timescale of the graph was chosen to capture the evolution
from the coherent to the thermodynamic regime, which is reflected
by the decreasing envelopes of magnetization and $m_F=0$
population. In order to identify further dephasing and decoherence
timescales we will now concentrate on the evolution of the $m_F=0$
population, which is not visibly affected by quadratic Zeeman
dephasing. The oscillation of both normal and condensed $m_F=0$
population is damped and initially tends towards its mean value of
0.25. This damping could be explained by random or spatial
dephasing e.g.~due to magnetic field gradients~\cite{Higbie2005a},
or by breakup into small dynamical spin domains due to a dynamical
instability~\cite{Mur2005a}.  On this timescale, dephasing should
at least in principle be reversible. In contrast, the drifting
apart of normal and condensed $m_F=0$ population at $\approx
15\,\mathrm{ms}$ marks the crossover to the thermodynamic regime
on longer timescales. In this regime, the system tends to restore
thermal equilibrium meaning in particular equal population
(i.e.~0.33) of all spin components in the thermal
cloud~\cite{Lewandowski2003a,Erhard2004a}.  The main process here
is redistribution of atoms between condensed fraction and thermal
cloud within each spin component, leading to the diverging trend
in $m_F=0$ populations observed.

\subsection{Population evolution}

In order to observe the effect of spin interaction beyond the
single particle level, we investigated the resulting population
evolution of the intermediate Ramsey state $\inistat$ directly,
i.e.~in a sequence of a $\pi /2$-pulse and a variable delay before
detection.  In contrast to the above Ramsey experiment this
sequence is not sensitive to the relative phases of the spinor
components but gives complementary information on the interaction
driven population evolution.

The quadratic Zeeman dominated regime at high magnetic fields is
particularly interesting for the investigation of coherence in
interaction driven spin dynamics. Our theoretical results
Eq.~(\ref{eq:popana}) predict that the quadratic Zeeman phase
determines the direction of spin mixing dynamics,
demonstrating the phase sensitivity of the process.

\begin{figure}[t]
\includegraphics{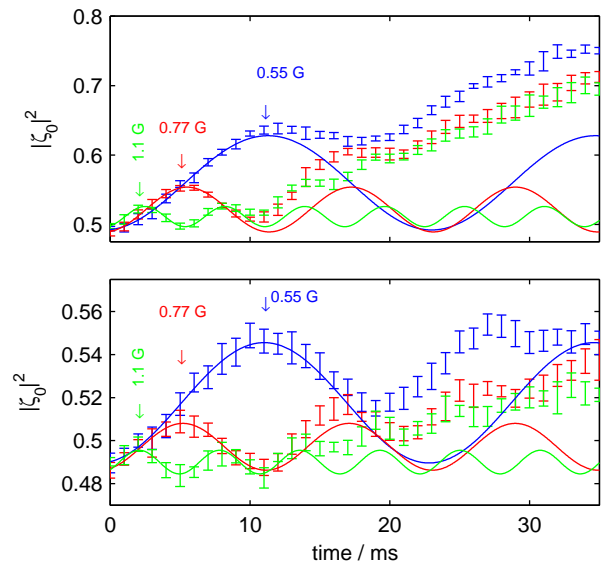}
\caption{\label{fig:Oscillations} Evolution of the $m_F=0$
  population in the condensate fraction for ensembles containing $\approx
  10^6$ atoms with $25\ldots 30\,\%$ condensate fraction (upper figure) and
  $\approx 4\cdot 10^5$ atoms with $75\ldots 80\,\%$ condensate fraction (lower figure),
  prepared in the state $\zeta_{\frac{\pi }{2}} $. The data points
  correspond to several measurements binned in 1\,ms intervals and error bars
  representing $\pm$ the standard deviation. The blue/red/green curves
  correspond to measurements at different offset fields with quadratic Zeeman
  periods of $\pi /q \approx 23/11.7/5.7\,\mathrm{ms}$ respectively. The lines
  correspond to the analytic solution~(\ref{eq:popana}), with parameters $q$
  calculated from the magnetic offset field and $k$ obtained from fitting to
  the first $10\,\mathrm{ms}$ of data. Additional fit parameters account for
  detection errors (y-offset) and evolution during time of flight (x-offset).
}
\end{figure}

The results shown in Fig.~\ref{fig:Oscillations} clearly confirm the
coherence of spin dynamics in spinor Bose-Einstein
condensates~\footnote{It is interesting to note that the initially
  prepared spinor corresponds to the ferromagnetic ground state (a
  stretched state with $|\bm F|^2=1$) while the fully dephased spinor
  corresponds to an anti-ferromagnetic ground state ($|\bm F|^2=0$) of
  the spin interaction Hamiltonian $H_\text{mf}$. In the quadratic
  Zeeman regime, the spinor is thus periodically transformed between
  these two extremal states, which both show no spin dynamics.  A
  comparison with the Ramsey experiment shows that the extrema of the
  $m_F=0$ population oscillations indeed coincide with the nodes and
  antinodes of the Ramsey amplitude envelope.}.
Furthermore the positive initial slope of the population evolution
of the $m_F=0$ state confirms the ferromagnetic behavior of
$^{87}$Rb in the F=1 hyperfine
manifold~\cite{Schmaljohann2004a,Chang2004a}.

\begin{figure}[t]
\includegraphics{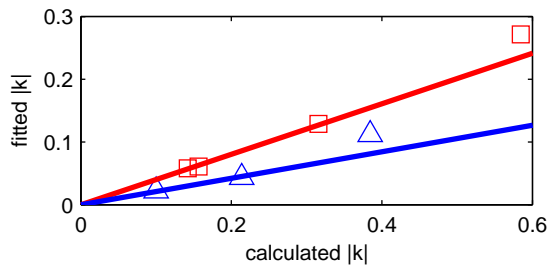}
\caption{\label{fig:kvalue} Fit values of the $|\zeta_0|^2$
  oscillation amplitude parameter $k$ of Eq.~(\ref{eq:popanaa}) versus the
  theoretical expectation for measurements with $25\ldots 30\,\%$ condensate
  fraction (squares) and with $75\ldots 80\,\%$ condensate fraction
  (triangles). The lines correspond to linear fits excluding the highest $k$
  data points.}
\end{figure}

Comparison with theory (Eq.~\ref{eq:popana}) shows an excellent
agreement during the first $10\,\mathrm{ms}$ of evolution, in
which a coherent behavior is expected considering the results of
the Ramsey experiment. The crossover to the thermodynamic regime
at later times goes along with damping of the population
oscillations and a transfer of particles between the condensate
fraction and normal component. The thermodynamic nature of this
process also shows up as a temperature dependence with longer
decoherence times for the lower temperature data.  It is important
to note, that the damping of the oscillations observed in this
case is fundamentally different from the Ramsey experiment. The
Ramsey experiment is sensitive to both spatial dephasing and
thermal decoherence, while the ''single $\pi/2$-pulse experiment''
is only affected by thermal decoherence. The relatively long
lasting $m_F=0$ oscillations in this experiment thus corroborate
our interpretation that the faster decay in the
 Ramsey experiment is not of thermal origin, but probably due
to magnetic field dephasing or more complex effects~\cite{Mur2005a}.

The amplitude of the population oscillations is proportional to
the spin dependent interactions via the parameter $k$ in
Eq.~(\ref{eq:popana}). In order to compare the experimental
results to our theoretical prediction, we plot the fit value
(fitting the analytic solution to the first 10\,ms of data) versus
the calculated value (using the scattering lengths
$a_{F=2}=100.4(1)\,a_0$ and
$a_{F=0}=101.8(2)\,a_0$~\cite{Kempen2002a} and the Thomas-Fermi
approximation to obtain $\langle n \rangle$) in
Fig.~\ref{fig:kvalue}. The fitted values indicate a lower
effective interaction parameter of 0.2 to 0.4 times the calculated
value and surprisingly show a clear temperature dependence.
Possible explanations include the breakdown of the Thomas-Fermi
approximation in the tightly confined vertical direction, trap
anharmonicities or small scale domain formation and require
further investigations.

In conclusion we have analyzed the interplay of external magnetic
fields and interatomic interactions on the system evolution,
giving analytic expressions in comparison to experimental data.
The experimental data at finite temperature shows various
dephasing/decoherence mechanisms, which we were able to separate
using the theoretical model and adapted experimental sequences.
These investigations pave the way to use spinor condensates for
general studies of dephasing and decoherence phenomena in
multi-component systems.

This work was funded in part by Deutsche Forschungsgemeinschaft
(DFG) within SPP 1116. R.W. acknowledges gratefully support from
the Atomics program of the Landesstiftung Baden-W\"urttemberg.
P.N. acknowledges support from the German AvH foundation and from
Junior Fellowship F/05/011 of the KULeuven Research Council.

\end{document}